# One size does not fit all: the complex relationship between biodiversity and psychological well-being

Assaf Shwartz*, Maya Tzunz, Lee Gafter and Agathe Colléony

Human and Biodiversity Research Lab, Faculty of Architecture and Town Planning, Technion – Israel Institute of Technology, Haifa, 32000 Israel

*Corresponding author: Assaf Shwartz (shwartza@technion.ac.il)
Maya Tzunz: mayatzunz@gmail.com
Lee Gafter: leegafter@gmail.com
Agathe Colléony: agathe.colleony@gmail.com

# Abstract

Enhancing urban biodiversity is increasingly advanced as a nature-based solution that can help align public health and biodiversity conservation agendas. Yet, research on the relationship between biodiversity and psychological well-being provides inconsistent results. The goal of this interdisciplinary research was to understand how components of psychological well-being of green space users relate to species richness and abundance. Additionally, we investigated how key characteristics that shape the way people interact with nature, affinity towards nature and ecological knowledge, moderate the well-being biodiversity relationship. We sampled bird, butterfly and plant in 24 urban gardens in Israel and distributed 600 close-ended questionnaires in-situ to measure psychological well-being, nature-relatedness, ecological knowledge, perceived species richness and demographics. Components of psychological well-being were mostly associated with perceived species richness and to lesser extent with actual species richness and abundance for all taxa. Nature-relatedness moderated these relationships. Respondents with high nature-relatedness demonstrated positive well-being-richness relationships, while those with intermediate, or low, nature-relatedness showed no, or even negative relationships, respectively. Opposite relationships were recorded for bird abundance, i.e., negative versus positive well-being-abundance relationship for individuals with high or low nature-relatedness, respectively. Overall, individuals demonstrated poor ecological knowledge of species and this variable moderated the relationships between well-being components and perceived butterfly richness and bird abundance. Our results demonstrate that one-size-does-not-fit-all when considering the relationship between psychological well-being and biodiversity and that affinity to nature is a key moderator for this relationship.

# Introduction

Urbanization deletes and degrades natural habitats and threatens many species (Antrop, 2004; McKinney, 2002) and increases the distance between residential and natural areas, progressively alienating people from the experience of nature (Turner et al., 2004). This is profoundly concerning given the mounting empirical evidence demonstrating that interactions with nature is important for human well-being and for maintaining affinity towards and care for the natural world (Colléony et al., 2020; Marselle et al., 2019; Mayer & Frantz, 2004). Slightly more than half of the world population currently lives in cities today, and this proportion is expected to increase to 68% by 2050 (United Nations, 2018). Therefore, the deleterious effects of individuals' alienation from nature on their well-being are likely to aggravate if nothing is done. Restoring or enhancing nature interactions in cities is important for safeguarding human and biodiversity health and conserving biodiversity in cities is often suggested as a strategy to support both biodiversity and human well-being (Shwartz, Turbé, Julliard, et al., 2014). However, understanding of the relationship between biodiversity and psychological well-being remains poor (reviewed by Marselle et al., 2019) and strategies benefiting people's health and well-being may not align with those targeting biodiversity conservation (Clayton et al., 2017; Colléony & Shwartz, 2019). It is therefore key to understand the relationships between biodiversity and psychological well-being and the variables that determine it for guiding urban planning and conservation practices that jointly benefits people and biodiversity.

Mounting empirical evidence suggests that the relationship between biodiversity and well-being is not linear, and even shows inconsistent patterns (Pett et al., 2016; Shanahan et al., 2015). First, there were more studies showing no significant results than studies providing evidence for a relationship between biodiversity and well-being (Marselle et al., 2019). Second, this relationship can vary across taxa and cultural and urban contexts. Studies found a positive associations between bird species richness and psychological well-being (Cameron et al., 2020; Dallimer et al., 2012; Fuller et al., 2007; Luck et al., 2011; Wolf et al., 2017; Wood et al., 2018) and similar patters were also recorded for the diversity of birdsongs (Ferraro et al., 2020; Hedblom et al., 2014). Butterfly species richness was not associated with any psychological well-being measure (Dallimer et al., 2012; Fuller et al., 2007). Plant species richness was positively associated with psychological well-being in some studies (Fuller et al., 2007; Wolf et al., 2017; Wood et al., 2018), whereas this relationship was negative, unimodal or showed no pattern in others (Dallimer et al., 2012; Southon et al., 2018). For instance, no association was found between vegetation complexity and mental health (Shanahan et al., 2016), while high density of trees around people's homes (but not richness) was related to lower rate antidepressant prescriptions (Marselle et al., 2020). A unimodal relationship between plant richness and stress recovery was found in an experiment conducted in urban park in Zurich Switzerland (Lindemann-Matthies & Matthies, 2018). Two additional experiments demonstrated that people did not perceive changes in species diversity (Shwartz, Turbé, Simon, et al., 2014; Southon et al., 2018). Finally, studies also showed that individuals can sometimes respond to the biodiversity they perceive to be present, rather than the biodiversity actually present in a greenspace (Dallimer et al., 2012; Southon et al., 2018).

Several factors can explain these complex and conflicting relationships. First, context may influence findings. There are cultural differences in human nature relationships (Colléony et al., 2019), but most studies on the biodiversity and well-being relationship were conducted in Western Europe and Anglo-Saxon countries (Marselle et al., 2019; Pett et al., 2016). Urban context can also affect this relationship, as complex environments such as riparian areas may induce more stress on some people than managed green spaces like public gardens and parks (Clayton et al., 2017). To date, much of the research that explored the relationships between biodiversity and well-being has focused on large green spaces or the general environment around participants homes (Marselle et al., 2019; Shwartz, Turbé, Simon, et al., 2014). Second, most studies, to date, have focused on species richness, habitat or ecosystem diversity as indicators for biodiversity (Marselle et al., 2019), while other measures such as abundance can also be influential. Finally, characteristics of individuals can also moderate the biodiversity–well-being relationship and in some cases demographic variables appeared to be more important in explaining well-being than biological ones (Luck et al., 2011; Southon et al., 2018). Yet, a recent review has shown that only four studies explored moderating variables such as, gender, age and socio-economic status in their analysis (Marselle et al., 2019).

Individuals develop a sense of connection with the natural world when experiencing nature, especially during childhood (Chawla, 2020). Consistently, individuals with a strong bond with the natural world are more prone to interact with and benefit from nature and positive associations were found between nature relatedness, nature interactions and well-being (Colléony et al., 2020; Martin et al., 2020). Spending time in nature provides individuals the opportunity to distract attention toward pleasant stimuli, rest, reflect and restore their self, as stated by the attention restoration theory (Kaplan & Kaplan, 1989). In accordance, studies have shown that nature relatedness is positively associated with the time people spend in nature and outdoors (e.g., Nisbet et al., 2009), and with lower level of state and trait cognitive anxiety (e.g., Martyn & Brymer, 2016). Thus, individual differences in the levels of nature relatedness might affect the well-being benefits derived from nature interaction (Lawton et al., 2017; Nisbet et al., 2011). Yet, to the best of our knowledge only one attempt was made to explore the relationship between biodiversity, nature relatedness and well-being benefits. This was tested in a meadows experiment which demonstrated a positive association between perceived species richness and connection to nature (Southon et al., 2018).

The relationship between biodiversity and well-being can also be moderated by people's ecological knowledge, and specifically their level of familiarity with a given species, which was found to be associated with attitudes toward species (Lindemann-Matthies, 2005; Luna et al., 2019; Sweet et al., 2020). The ability to name species (hereafter species identification skills) is a basic component of people's relationship with nature (Kai et al., 2014) and therefore likely to affect the benefits individuals retrieve from biodiversity when experiencing it. Few studies have shown that individuals have poor species identification skills and suggested that this can impact their ability to perceive biodiversity and benefit from it (reviewed by Pett et al., 2016). For instance, Southon et al., (2018) have shown that respondents with higher eco-centricity (also comprising of plant identification skills) were more accurate in estimating species richness of restored meadows in Southern England. As they also found a correlation between perceived species richness and well-being benefits, they postulate that people with eco-centric traits are more likely to benefit from biodiversity. Species identification skills are therefore likely to affect the benefits individuals retrieve from nature when experiencing it. Yet, biodiversity and psychological well-being are complex and multifaceted concepts, and much research is still required to understand the relationships between them and the moderating effects that influence this relationship (Marselle et al., 2019; Shanahan et al., 2015). Such knowledge is key to align the agendas of public health and biodiversity conservation.

In this study, we aimed to bridge this important knowledge gap, by exploring the relationship between biodiversity and psychological well-being, and the effect of two moderating variables, species identification skills and nature relatedness, while accounting for demographic variables. Building on and expanding previous studies (Dallimer et al., 2012; Fuller et al., 2007), we conducted social and ecological (birds, butterflies and plants) surveys in small public urban gardens to measure components of psychological well-being and various biodiversity indicators. The study was conducted in Israel, where such interdisciplinary study was never performed yet, and where people's level of nature relatedness and species identification skills was found to be lower than European countries (Colléony et al., 2019). Our goal was to understand how components of psychological well-being are related to actual species richness, abundance and perceived species richness for visitors of small public recreational gardens that people frequent in their daily lives. We were also interested to understand how nature relatedness and species identification skills can moderate these relationships.

**Methods**

*Ecological survey*

The study was conducted in Netanya, a coastal city of approximately 200,000 residents in Israel (CBS, 2017). We selected 24 small public recreational gardens, located throughout the city. Gardens were carefully selected to meet certain criteria: (1) size – we selected gardens ranging from 0.5 to 3.1 ha; (2) urban context – all gardens were located within residential areas and were planned and managed for recreation activity of the dwellers in the neighborhood; (3) socioeconomic context – gardens were selected to represent the different socioeconomic levels across the city; and (4) avoiding sea view – to minimize the effect of blue space on well-being, we decided to exclude gardens with a view of the sea. The selection process began by identifying all the green spaces in Netanya, using an aerial photograph and the city's land use map (updated to 2014), supplied by the municipality planning and GIS department. To obtain more information on the locations and characteristics of Netanya's green spaces we interviewed the head of the parks and recreation department. This process yielded a preliminary list of 96 green spaces, sized 0.08 to 5.5 hectares. We visited the 96 locations during winter 2015 and collected spatial data including land cover and boundaries, the existence of physical elements such as a playground, pond or unique facilities and other exceptional qualities (e.g., old trees, steep slopes or odd garden shapes). We then calculated the area of each garden using GIS and looked at the socioeconomic status of the neighborhood in which it was located. The Israeli central statistics bureau characterizes each neighborhood to a socio-economic cluster ranging from 0-20 based on 16 variables that include education, employment, quality of life and demographics. In Netanya the scores ranged from 5-16 and our gardens were selected to represent this range as much as possible.

We surveyed three taxonomic groups (birds, butterflies and plants) in each garden during spring (April – June) 2015. Sampling of all groups was standardized to an area of one hectare, by limiting data collection to a selected 1 ha quadrat in larger gardens and including green and grey spaces adjacent to the small gardens (N=14). For gardens larger than 1 ha the quadrat was located randomly in the garden, but in a way that ensured that all quadrat was located within the garden. Birds were surveyed during breeding season (April-June 2015), using point counts, on eight visits at each garden. We aimed to visit each garden once every 7-10 days but given some practical and weather constrains this objective was not fully met. Every bird seen or heard up to 50 meters from the sampling point (in the center of the abovementioned quadrat) was recorded during 10-minutes visits. Visits were held early in the morning, from half an hour before sunrise and up to three hours after sunrise, depending on the weather conditions. Each record included the species name, number of individuals, location, and distance from observer as measured using a laser range finder.

Diurnal butterflies (Lepidoptera sp.) were sampled eight times using the quadrat method (following Shwartz, Muratet, et al., 2013) from April to June 2015 on sunny days with temperature above 18 degrees $C^0$, typically from 0900 to 1700. We used the quadrat method as it was more suitable than normal transect sampling for the gardens that were relatively individual- and species-poor. Thus, butterflies were sampled in the 1 ha quadrat in which we directionally strolled (never returning back) for 15 min recording any butterfly in sight. When needed, in order to verify identification, we captured and immediately released butterflies with a sweeping net. All butterflies were identified at the species level except small whites, which were grouped at the genus level (i.e., Pieris). Plant survey was held at the beginning of April, during the peak of the blooming period, and was led by skilled botanist H. Leshner. During the botanic survey we recorded plant species in the gardens, within the quadrat. Given that people notice mainly woody and flowering species (Fuller et al., 2007; Lindemann-Matthies et al., 2010; Shwartz, Turbé, Simon, et al., 2014), we recorded all woody species and flowering species, meaning herbaceous species were only recorded when in bloom. We strolled along each garden for approximately 30 minutes and for each species recorded its name, whether it was native or not, whether it was flowering or not in the time of visit (peak flowering season).

Bird and butterfly species richness were calculated as the cumulative number of species sampled in the eight visits; abundance was derived by averaging the number of individuals observed in the eight visits. For plants, we calculated species richness of flowering species and woody species (i.e., trees and bushes). We used GIS to measure the covers of woody species (trees and shrubs), lawn, bare ground and grey infrastructures (e.g., buildings or roads) using on high-resolution orthophoto and verified these measures while visiting the gardens using GIS-Collector. Following Shwartz et al. (2013), we used the land cover proportions to calculate the Shannon index of habitat diversity (SHDI) as follows: $SHDI = \sum_{i=1}^{m} Pi * lnPi$, with Pi standing for the proportion of any specific land cover within the garden.

*Social survey*

During Spring 2015, we surveyed 600 adults in the 24 gardens. We measured *psychological well-being* derived from nature visits with Fuller et al. (2007) scale, a scale derived from the attention restoration theory and concepts of emotional attachment to, and personal identity gained from, the greenspace. We measured *nature relatedness* [NR] with the 6-item Nature Relatedness Scale (Nisbet & Zelenski, 2013). Respondents were asked to report the extent to which they agree on each item using five-point Likert scales (from strongly disagree to strongly agree). Following Fuller et al. (2007), we conducted a factor analysis to identify groups of statements measuring a single component of *psychological well-being* (Table A1). After confirming satisfactory internal consistency, we averaged scores of items of each subscale to derive three single measures of psychological well-being, *attention restoration* (Cronbach alpha, α = 0.71), *attachment* (α = 0.83) and *sense of identity and continuity with past* (α = 0.79). We also averaged scores of items of nature relatedness to derive a single measure of *nature relatedness* (α = 0.77).

*Perceived species richness* was measured following Fuller et al. (2007). We asked respondents to estimate the number of species of each taxonomic group, from a given list of ranges, based on preliminary observations in the garden and based on in site preliminary ecological survey. A five-point Likert scale was used to represent the different ranges of species richness for each taxa: (i) for birds – 0, 1-6, 7-12, 13-18, 19+; (ii) for butterflies – 0, 1-3, 4-6, 7-9, 10+; (iii) for trees – 2, 3-6, 7-10, 11-14, 15+; and (iv) for flowers – 0-5, 6-10, 11-15, 16-20, 21+. We also measured species identification skills, as a proxy for *ecological knowledge* (following Dallimer et al. 2012). We created a plate displaying 12 pictures of common species (birds, butterflies and plants; see Figure A1) and asked respondents to name each of those species. We interviewed ornithologists, botanists and lepidopterologists with experience in outreach activities to select ten native and two non-native species (one bird and one flowering plant species) that are widespread, well recognized and expected to be found in the study sites. Each respondent was given one score of species identification skills corresponding to the number of genera they named correctly. In light of answers' level of specificity, we decided to accept answers naming the correct genus rather than solely the precise full species name. For instance, an identification of the butterfly *Vanessa cardui* just by its genus – Vanessa – was accepted as correct. Thus, our measure was less conservative than the one developed by Dallimer et al. (2012), since we accepted genera as correct answers and given that although the species selected appeared in most gardens, the selection of species was not based on *in situ* ecological surveys.

Finally, each participant was also asked to provide their *gender*, *age*, *perceived income* (using a five-point Likert scale, from much lower to much higher than Israel average wage), *level of education* indicated by years of schooling, urbanization of place of *childhood residence* (large, medium, small sized city, a village or community settlement [Moshav or Kibbutz]). The survey was distributed in Hebrew. Each scale was first translated into Hebrew by one researcher (MT) then back translated into English by another researcher (AS) to verify that translation retained the original meanings of the original NR scale. The survey was piloted with 13 individuals before distributing to garden visitors.

*Statistical analyses*

Spatial analyses were conducted using ArcGIS 10.2 and statistical analyses using R 3.6.0 (R Core Team, 2013). We built three generalized linear models with Poisson error distribution to explore how perceived species richness relates to observed richness of the corresponding taxa (birds, plants, butterflies), nature relatedness and species identification skills. We then built 12 similar linear models to relate the three components of psychological well-being and biodiversity indicators for each taxon separately, as observed richness was significantly correlated between taxa. In each model, we explored how the component of psychological well-being relates to perceived and observed richness, observed abundance (for birds and butterflies), nature relatedness and species identification skills. Gardens' size and sociodemographic variables were added to all models to account for their potential effect (see Luck et al., 2011).

Interactions between observed and perceived species richness (for all taxa), abundance (for birds and butterflies) and nature relatedness and species identification skills were added to test for their moderating effect. Interactions were plotted using the plot_model in R package sjPlot (Lüdecke et al., 2021). The function plots interactions of continuous terms by dividing them into groups (low, medium, high). We also estimated the share of respondents in each group by dividing the range of scores to three groups and counting the number of respondents in each group. We checked the normality assumption by plotting the residuals and checked for multicollinearity using variance inflation factors (car package in R). We conducted stepwise model selection based on the Akaike Information Criterion; stepwise deletion was carried out based on non-significant p-values ($p>0.05$), with largest p-values and interactions removed first.

*Ethics Statement*

Permission for this study was granted by the Technion Social and Behavioral Sciences Institutional Review Board (approval numbers 2015-123), and the research was performed in accordance with the board's relevant guidelines and regulations.

**Results**

We recorded 32 bird species, 13 butterfly species and 285 plant species (3 to 49 per garden) in total. On average, abundance per visit was 42.35±2.06 (AV±SE, hereafter) for birds and 6.55±0.68 for butterflies. Respondents were mostly women (60%), of 40±0.57 years old on average. *Nature relatedness* (NR) was on average 3.70±0.02, with 8% of respondents with low NR, 51% with medium NR and 41% with high NR. *Species identification skills* were very low (2.21±0.07 out of 12), most respondents (90%) correctly identify 4 species at most, and 62% of the respondents could only correctly identify one or two species (Fig. 1).

Individuals with high NR reported higher perceived species richness for plants than individuals with lower NR scores (Table 1; Fig. 2a). Garden size was positively associated with perceived butterfly and plant richness (Table 1). The relationship between perceived and observed species richness for butterflies was positive for individuals of high NR, and negative for individuals of low NR, although this interaction effect was only marginally significant (Table 1; Fig. 2b). There was no relationship between species identification skills and perceived species richness for any of the three studied taxa (Table 1).

Generally, psychological well-being measures were positively associated with garden size (Table 2). *Attachment* and *sense of identity and continuity with past* were negatively associated with observed bird richness (Table 2). Nature relatedness moderated the associations between perceived richness of plants, butterflies and woody cover and different psychological well-being measures (Fig. 3; Table 2). Patterns for all these interactions were similar and demonstrated positive, no trend and negative association for participants with high, medium and low nature relatedness respectively (Fig 3a, c-h). For instance, as perceived species richness of butterflies increased, *attention restoration* increased for individuals of high NR, but decreased for those of low NR (Fig. 3a). Similar moderating effect was also recorded for the association between butterfly abundance and *sense of identify and continuity with past* (Fig. 3b). In contrast, opposite associations were recorded for bird abundance. As bird abundance increased, *attention restoration* and *sense of identity and continuity with past* increased for individuals with low NR but decreased for individuals with high NR (Table 2; Fig. 3i-j). Furthermore, as perceived species richness of butterflies increased, *attachment* and *sense of identity and continuity with past* increased for individuals of high *species identification skills*, but decreased for those of low *species identification skills*; the negative association between bird abundance and *attachment* was stronger for individuals of high *species identification skills* than those of low *species identification skills* (Table 2; Fig. 4).

**Discussion**

Conserving biodiversity in cities is often advocated as a win-win nature-based solution to mitigate the detrimental impacts of urbanization on both people and biodiversity conservation (Dearborn & Kark, 2010; Miller, 2005; Shwartz, Turbé, Julliard, et al., 2014). However, although many studies have established the relationship between nature and health and well-being, our understanding of the biodiversity – well-being relationship remains poor and this relationship is not as straightforward as expected (Marselle et al., 2019; Pett et al., 2016). In this study, we show that, consistently with previous studies, psychological well-being is related to perceived rather than actual species richness. More importantly, we demonstrate that the biodiversity – well-being relationship can be moderated by individual's characteristics, such as nature relatedness and species identification skills. While individuals with high nature relatedness and those with high species identification skills benefited from gardens that were or that they perceived to be biodiversity rich, benefits decreased for individuals with low nature relatedness. Thus, one size may not fit all and enhancing biodiversity in urban green spaces may not necessarily directly maximize well-being for all visitors of green spaces. The moderating effects of nature relatedness and to lesser extent species identification skills, highlight the importance of diversifying urban green space types, so they will provide inclusive nature experiences for a range people and not only those who are well connected to nature and benefit from biodiversity. This can ultimately facilitate a range of positive interactions with nature, stronger connection to nature and knowledge that then in turn can influence the benefits people gain from visiting biodiverse green spaces.

Nature experiences provide a wide range of health and well-being benefits to people and promote a sense of affinity toward the natural world (Clayton, 2012; Keniger et al., 2013). Several studies have argued and showed that nowadays, the direct experience of nature of city dwellers is reduced (Clayton et al., 2017; Soga & Gaston, 2016, 2018), although retrospective studies have shown that contact with nature remained comparable for adults (Novotný et al., 2020) and children (Oh et al., 2020). A decrease in direct interactions thus threatens the delivery of health and well-being benefits from nature experiences. Individuals experiencing nature less directly, and thus with lower nature relatedness, are likely to perceive ecologically rich areas more messy and less attractive (Nassauer, 1995). Alternatively, they may gain the greatest well-being benefits from experiencing low biodiversity gardens because of the novelty and excitement. This can potentially explain the negative relationship that we found between perceived species richness and psychological well-being for some individuals. In contrast, individuals with high nature relatedness are more likely to have high quality nature interactions and the associated well-being benefits (Colléony et al., 2020; Prévot et al., 2018). Consistently, previous studies have showed that perceptions of biodiversity influence perceived restorativeness, which subsequently influences human well-being (Fisher et al., 2021; Marselle et al., 2016).

Interestingly, psychological well-being measures were negatively associated with bird abundance for individuals of high nature relatedness or species identification skills, while this relationship was positive for individuals of low nature relatedness. A possible explanation for this opposite trend is the higher abundance of synanthropic species such as the domestic pigeon, or invasive alien ones such as the common myna, in comparison to other species. Individuals more connected or knowledgeable about biodiversity may be more aware of the disturbances caused by these species (e.g., displacement of native species, harassment of passerby), which could alter the restorative benefits of their experience in nature. It is therefore possible that different species, or components of nature, influence differently individual well-being. Colorful or charismatic species are usually preferred compared to less colorful or more common species (Shwartz, Cheval, et al., 2013), and these preferences may affect restorative benefits of nature interactions. Experimental studies comparing the relationship between different assemblage of bird species (richness and abundance) and psychological well-being (e.g., with images, movies or virtual reality experiments) could help shed light on this phenomenon.

Noteworthy, our findings suggest that individuals with high nature relatedness perceive species richness more accurately than those with low nature relatedness. Individuals with high affinity towards nature are more inclined to directly interact with nature than those with low affinity towards nature (Colléony et al., 2020). Paying attention to and showing interest towards the natural world may thus help to detect differences between species. In contrast, we suspect that individuals with low nature relatedness underestimate species richness by not detecting differences between species. Interestingly, there was no relationship between species identification skills and perceived richness. Our sample of respondents showed very little skills at identifying common species, which may explain that we found no relationship. Previous studies have noted similar low species identification skills, although this varied across countries, urban contexts and taxonomic groups (Bashan et al., 2020; Colléony et al., 2019; Dallimer et al., 2012; Southon et al., 2018). However, the ability to adequately compare species identification skills of our respondents to those of Dallimer et al. (2012) is limited, given that methodologies and contexts slightly differed. In particular, we focused on public gardens, a place accessible to a variety of different types of individuals, and we acknowledge that results could differ if focusing on less managed natural settings such as riparian areas.

Nevertheless, we feel that further investigation of species identification skills in various countries and contexts is important. This is because ignorance can lead to indifference and since naming species can be considered as the basic component of acquaintance with nature (Kai et al., 2014). It is thus important to monitor how different species identification skills and other types of ecological knowledge vary as our world is becoming more and more urbanized, as well as the consequences of these changes. A recent study has already highlighted some cultural differences in species identification skills between Israel and two European countries (Colléony et al., 2019) and another study showed that biodiversity-well-being relationship can vary between countries and cities (Fischer et al., 2018). However, it is also plausible that people feel familiar with a given species or benefit from species diversity without being able to name them correctly. Further research is still needed to understand the extent to which knowing species can influence the well-being biodiversity relationship. Altogether results of this research highlight the importance of exploring moderating variables and future research could also benefit from exploring various variables as moderators for instance using structural equation modeling. Innovatively considering the potential moderating role of nature relatedness and ecological knowledge in the relationship between psychological well-being and biodiversity, this study demonstrates that one size does not fit all when considering psychological well-being benefits in relation to biodiversity.

For the majority (51%) of the respondents with medium level of nature relatedness perceived and actual level of species diversity did not influence the biodiversity-well-being relationship, while respondents with high (41%) and low (8%) nature relatedness demonstrated opposite positive and negative relations respectively. Therefore, it is important to underline that efforts to restore or enhance biodiversity in urban green spaces can have various consequences for different people and in some circumstances biodiverse gardens may even deter visits, reduce nature interactions and well-being benefits. We recommend urban planners and conservation practitioners to focus their efforts not only on enhancing biodiversity but also on promoting emotional connection to and knowledge about nature. Given that different individuals in different cultures and circumstances already differ in their nature relatedness and ecological knowledge (Bashan et al., 2020; Colléony et al., 2019), we think it is important to also diversify green spaces, varying in their level of biodiversity to enable diverse and inclusive nature experiences for various people. This is also supported by a recent study which demonstrated that individuals who experienced different types of green spaces (e.g., in terms of level of species richness) reported higher life satisfaction (Chang et al., 2020). However, it is important to note results are based on a case study, therefore calling for caution regarding any generalizability of the findings.

Biophilic designs can be used to raise individuals' attention towards biodiversity and inform them on species names (e.g. signs with name of plants next to the plant, or plate with names of most common birds observed in the park) (Beatley, 2010; Shwartz, 2018). Citizen science can also help enhance species identification skills and nature relatedness (Schuttler et al., 2018). Furthermore, our results emphasized that promoting emotional connection to nature is even more important than informing the public about biodiversity, as the effect of nature relatedness was stronger and more consistent across taxa and well-being indicators than the effect of ecological knowledge. Not all individuals need to be knowledgeable about biodiversity but connecting emotionally with nature will benefit their well-being and, potentially, foster commitment to protect the natural world (Chawla, 2020). Providing more green spaces may thus not be sufficient and landscape planners should design green spaces that enable meaningful nature interactions that can in turn foster affinity towards nature. Subtle stimuli or short messages such as 'cues to experience nature' can be used to bring people close to nature and promote affinity towards nature and high quality nature interactions (Colléony et al., 2020). Further effort should explore other means to promote emotional connection to nature.

**Acknowledgements**:

The research was funded by the European Union (ERC starting grant, Niche4NBS, project number 852633). Views and opinions expressed are, however, those of the author(s) only and do not necessarily reflect those of the European Union or the European Research Council Executive Agency. Neither the European Union nor the granting authority can be held responsible for them. The research was also supported by the Municipality of Netanya.

**Data Availability**

The data will be archived via the Dryad Digital Repository or similar.

**Supporting information**

Species presented to the respondents (Fig. A1) and factor loadings for the well-being components and their internal validity (Table A1) are provided in Supporting Information.

**References**

Antrop, M. (2004). Landscape change and the urbanization process in Europe. *Landscape and Urban Planning*, *67*(1–4), 9–26. https://doi.org/10.1016/S0169-2046(03)00026-4

Bashan, D., Colléony, A., & Shwartz, A. (2020). Urban versus rural? The effects of residential status on connection to nature and species identification skills. *People and Nature*.

Beatley, T. (2010). *Biophilic cities* (Island Press).

Cameron, R. W. F., Brindley, P., Mears, M., McEwan, K., Ferguson, F., Sheffield, D., Jorgensen, A., Riley, J., Goodrick, J., Ballard, L., & Richardson, M. (2020). Where the wild things are! Do urban green spaces with greater avian biodiversity promote more positive emotions in humans? *Urban Ecosystems*, *23*(2), 301–317. https://doi.org/10.1007/s11252-020-00929-z

CBS. (2017). *Population Census—Central Bureau of Statistics*. Population Census - Central Bureau of Statistics. http://www.cbs.gov.il/census/census/main_mifkad08_e.html

Chang, C., Oh, R. R. Y., Nghiem, T. P. L., Zhang, Y., Tan, C. L. Y., Lin, B. B., Gaston, K. J., Fuller, R. A., & Carrasco, L. R. (2020). Life satisfaction linked to the diversity of nature experiences and nature views from the window. *Landscape and Urban Planning*, *202*, 103874. https://doi.org/10.1016/j.landurbplan.2020.103874

Chawla, L. (2020). Childhood nature connection and constructive hope: A review of research on connecting with nature and coping with environmental loss. *People and Nature*, *2*(3), 619–642. https://doi.org/10.1002/pan3.10128

Clayton, S. (2012). Environment and identity. In *Oxford Handbook of Environmental and Conservation Psychology* (Owford University Press, pp. 164–180). S. Clayton.

Clayton, S., Colléony, A., Conversy, P., Maclouf, E., Martin, L., Torres, A.-C., Truong, M.-X., & Prévot, A.-C. (2017). Transformation of experience: Toward a new relationship with nature. *Conservation Letters*, *10*(5), 645–651. https://doi.org/10.1111/conl.12337

Colléony, A., Levontin, L., & Shwartz, A. (2020). Promoting meaningful and positive nature interactions for visitors to green spaces. *Conservation Biology*, *34*(6), 1373–1382. https://doi.org/10.1111/cobi.13624

Colléony, A., & Shwartz, A. (2019). Beyond Assuming Co-Benefits in Nature-Based Solutions: A Human-Centered Approach to Optimize Social and Ecological Outcomes for Advancing Sustainable Urban Planning. *Sustainability*, *11*(18), 4924. https://doi.org/10.3390/su11184924

Colléony, A., White, R., & Shwartz, A. (2019). The influence of spending time outside on experience of nature and environmental attitudes. *Landscape and Urban Planning*, *187*, 96–104. https://doi.org/10.1016/j.landurbplan.2019.03.010

Dallimer, M., Irvine, K. N., Skinner, A. M. J., Davies, Z. G., Rouquette, J. R., Maltby, L. L., Warren, P. H., Armsworth, P. R., & Gaston, K. J. (2012). Biodiversity and the Feel-Good Factor: Understanding Associations between Self-Reported Human Well-being and Species Richness. *BioScience*, *62*(1), 47–55. https://doi.org/10.1525/bio.2012.62.1.9

Dearborn, D. C., & Kark, S. (2010). Motivations for Conserving Urban Biodiversity. *Conservation Biology*, *24*(2), 432–440. https://doi.org/10.1111/j.1523-1739.2009.01328.x

Ferraro, D. M., Miller, Z. D., Ferguson, L. A., Taff, B. D., Barber, J. R., Newman, P., & Francis, C. D. (2020). The phantom chorus: Birdsong boosts human well-being in protected areas. *Proceedings of the Royal Society B: Biological Sciences*, *287*(1941), 20201811. https://doi.org/10.1098/rspb.2020.1811

Fischer, L. K., Honold, J., Cvejić, R., Delshammar, T., Hilbert, S., Lafortezza, R., Nastran, M., Nielsen, A. B., Pintar, M., van der Jagt, A. P. N., & Kowarik, I. (2018). Beyond green: Broad support for biodiversity in multicultural European cities. *Global Environmental Change*, *49*, 35–45. https://doi.org/10.1016/j.gloenvcha.2018.02.001

Fisher, J. C., Irvine, K. N., Bicknell, J. E., Hayes, W. M., Fernandes, D., Mistry, J., & Davies, Z. G. (2021). Perceived biodiversity, sound, naturalness and safety enhance the restorative quality and wellbeing benefits of green and blue space in a neotropical city. *Science of The Total Environment*, *755*, 143095. https://doi.org/10.1016/j.scitotenv.2020.143095

Fuller, R. A., Irvine, K. N., Devine-Wright, P., Warren, P. H., & Gaston, K. J. (2007). Psychological benefits of greenspace increase with biodiversity. *Biology Letters*, *3*(4), 390–394. https://doi.org/10.1098/rsbl.2007.0149

Hedblom, M., Heyman, E., Antonsson, H., & Gunnarsson, B. (2014). Bird song diversity influences young people's appreciation of urban landscapes. *Urban Forestry & Urban Greening*, *13*(3), 469–474. https://doi.org/10.1016/j.ufug.2014.04.002

Kai, Z., Woan, T. S., Jie, L., Goodale, E., Kitajima, K., Bagchi, R., & Harrison, R. D. (2014). Shifting Baselines on a Tropical Forest Frontier: Extirpations Drive Declines in Local Ecological Knowledge. *PLOS ONE*, *9*(1), e86598. https://doi.org/10.1371/journal.pone.0086598

Kaplan, R., & Kaplan, S. (1989). *The Experience of Nature: A Psychological Perspective*. Cambridge University Press.

http://www.psichenatura.it/fileadmin/img/R._Kaplan__S._Kaplan_The_Experience_of_Nature__Introduction_.pdf

Keniger, L. E., Gaston, K. J., Irvine, K. N., & Fuller, R. A. (2013). What are the Benefits of Interacting with Nature? *International Journal of Environmental Research and Public Health*, *10*(3), 913–935. https://doi.org/10.3390/ijerph10030913

Lawton, E., Brymer, E., Clough, P., & Denovan, A. (2017). The Relationship between the Physical Activity Environment, Nature Relatedness, Anxiety, and the Psychological Well-being Benefits of Regular Exercisers. *Frontiers in Psychology*, *8*. https://doi.org/10.3389/fpsyg.2017.01058

Lindemann-Matthies, P. (2005). `Loveable' mammals and `lifeless' plants: How children's interest in common local organisms can be enhanced through observation of nature. *International Journal of Science Education*, *27*, 655–677. https://doi.org/10.1080/09500690500038116

Lindemann-Matthies, P., Junge, X., & Matthies, D. (2010). The influence of plant diversity on people's perception and aesthetic appreciation of grassland vegetation. *Biological Conservation*, *143*(1), 195–202. https://doi.org/10.1016/j.biocon.2009.10.003

Lindemann-Matthies, P., & Matthies, D. (2018). The influence of plant species richness on stress recovery of humans. *Web Ecology*, *18*(2), 121–128. https://doi.org/10.5194/we-18-121-2018

Luck, G. W., Davidson, P., Boxall, D., & Smallbone, L. (2011). Relations between urban bird and plant communities and human well-being and connection to nature. *Conservation Biology*, *25*(4), 816–826. https://doi.org/10.1111/j.1523-1739.2011.01685.x

Lüdecke, D., Bartel, A., Schwemmer, C., Powell, C., Djalovski, A., & Titz, J. (2021). *Package "sjPlot"* (2.8.9) [Computer software].

Luna, A., Edelaar, P., & Shwartz, A. (2019). Assessment of social perception of an invasive parakeet using a novel visual survey method. *NeoBiota*, *46*, 71–89.

Marselle, M. R., Bowler, D. E., Watzema, J., Eichenberg, D., Kirsten, T., & Bonn, A. (2020). Urban street tree biodiversity and antidepressant prescriptions. *Scientific Reports*, *10*(1), 22445. https://doi.org/10.1038/s41598-020-79924-5

Marselle, M. R., Irvine, K. N., Lorenzo-Arribas, A., & Warber, S. L. (2016). Does perceived restorativeness mediate the effects of perceived biodiversity and perceived naturalness on emotional well-being following group walks in nature? *Journal of Environmental Psychology*, *46*, 217–232. https://doi.org/10.1016/j.jenvp.2016.04.008

Marselle, M. R., Martens, D., Dallimer, M., & Irvine, K. N. (2019). Review of the Mental Health and Well-being Benefits of Biodiversity. In M. R. Marselle, J. Stadler, H. Korn, K. N. Irvine, & A. Bonn (Eds.), *Biodiversity and Health in the Face of Climate Change* (pp. 175–211). Springer International Publishing. https://doi.org/10.1007/978-3-030-02318-8_9

Martin, L., White, M. P., Hunt, A., Richardson, M., Pahl, S., & Burt, J. (2020). Nature contact, nature connectedness and associations with health, wellbeing and pro-environmental behaviours. *Journal of Environmental Psychology*, *68*, 101389. https://doi.org/10.1016/j.jenvp.2020.101389

Martyn, P., & Brymer, E. (2016). The relationship between nature relatedness and anxiety. *Journal of Health Psychology*, *21*(7), 1436–1445. https://doi.org/10.1177/1359105314555169

Mayer, F. S., & Frantz, C. M. (2004). The connectedness to nature scale: A measure of individuals' feeling in community with nature. *Journal of Environmental Psychology*, *24*(4), 503–515. https://doi.org/10.1016/j.jenvp.2004.10.001

McKinney, M. L. (2002). Urbanization, Biodiversity, and Conservation. *BioScience*, *52*(10), 883–890.

Miller, J. R. (2005). Biodiversity conservation and the extinction of experience. *Trends in Ecology & Evolution*, *20*(8), 430–434. https://doi.org/10.1016/j.tree.2005.05.013

Nassauer, J. I. (1995). Messy Ecosystems, Orderly Frames. *Landscape Journal*, *14*(2), 161–170. https://doi.org/10.3368/lj.14.2.161

Nisbet, E. K., & Zelenski, J. M. (2013). The NR-6: A new brief measure of nature relatedness. *Frontiers in Psychology*, *4*. https://doi.org/10.3389/fpsyg.2013.00813

Nisbet, E. K., Zelenski, J. M., & Murphy, S. A. (2009). The Nature Relatedness Scale: Linking Individuals' Connection With Nature to Environmental Concern and Behavior. *Environment and Behavior*, *41*(5), 715–740. https://doi.org/10.1177/0013916508318748

Nisbet, E. K., Zelenski, J. M., & Murphy, S. A. (2011). Happiness is in our Nature: Exploring Nature Relatedness as a Contributor to Subjective Well-Being. *Journal of Happiness Studies*, *12*(2), 303–322. https://doi.org/10.1007/s10902-010-9197-7

Novotný, P., Zimová, E., Mazouchová, A., & Šorgo, A. (2020). Are Children Actually Losing Contact with Nature, or Is It That Their Experiences Differ from Those of 120 years Ago? *Environment and Behavior*. https://doi.org/10.1177/0013916520937457

Oh, R. R. Y., Fielding, K. S., Carrasco, R. L., & Fuller, R. A. (2020). No evidence of an extinction of experience or emotional disconnect from nature in urban Singapore. *People and Nature*, *2*(4), 1196–1209. https://doi.org/10.1002/pan3.10148

Pett, T. J., Shwartz, A., Irvine, K. N., Dallimer, M., & Davies, Z. G. (2016). Unpacking the People–Biodiversity Paradox: A Conceptual Framework. *BioScience*, biw036. https://doi.org/10.1093/biosci/biw036

Prévot, A.-C., Cheval, H., Raymond, R., & Cosquer, A. (2018). Routine experiences of nature in cities can increase personal commitment toward biodiversity conservation. *Biological Conservation*, *226*, 1–8. https://doi.org/10.1016/j.biocon.2018.07.008

R Core Team. (2013). *R: A language and Environment for Statistical Computing*. R Foundation for Statistical Computing.

Schuttler, S. G., Sorensen, A. E., Jordan, R. C., Cooper, C., & Shwartz, A. (2018). Bridging the nature gap: Can citizen science reverse the extinction of experience? *Frontiers in Ecology and the Environment*, *0*(0). https://doi.org/10.1002/fee.1826

Shanahan, D. F., Bush, R., Gaston, K. J., Lin, B. B., Dean, J., Barber, E., & Fuller, R. A. (2016). Health Benefits from Nature Experiences Depend on Dose. *Scientific Reports*, *6*, 28551. https://doi.org/10.1038/srep28551

Shanahan, D. F., Fuller, R. A., Bush, R., Lin, B. B., & Gaston, K. J. (2015). The Health Benefits of Urban Nature: How Much Do We Need? *BioScience*, biv032. https://doi.org/10.1093/biosci/biv032

Shwartz, A. (2018). Designing Nature in Cities to Safeguard Meaningful Experiences of Biodiversity in an Urbanizing World. In *Urban biodiversity—From research to practice* (pp. 200–215). ROUTLEDGE & GSE Research. https://doi.org/info:doi/10.9774/GLEAF.9781315402581_14

Shwartz, A., Cheval, H., Simon, L., & Julliard, R. (2013). Virtual Garden Computer Program for use in Exploring the Elements of Biodiversity People Want in Cities. *Conservation Biology*, *27*(4), 876–886. https://doi.org/10.1111/cobi.12057

Shwartz, A., Muratet, A., Simon, L., & Julliard, R. (2013). Local and management variables outweigh landscape effects in enhancing the diversity of different taxa in a big metropolis. *Biological Conservation*, *157*, 285–292. https://doi.org/10.1016/j.biocon.2012.09.009

Shwartz, A., Turbé, A., Julliard, R., Simon, L., & Prévot, A.-C. (2014). Outstanding challenges for urban conservation research and action. *Global Environmental Change*, *28*, 39–49. https://doi.org/10.1016/j.gloenvcha.2014.06.002

Shwartz, A., Turbé, A., Simon, L., & Julliard, R. (2014). Enhancing urban biodiversity and its influence on city-dwellers: An experiment. *Biological Conservation*, *171*, 82–90. https://doi.org/10.1016/j.biocon.2014.01.009

Soga, M., & Gaston, K. J. (2016). Extinction of experience: The loss of human–nature interactions. *Frontiers in Ecology and the Environment*, *14*(2), 94–101. https://doi.org/10.1002/fee.1225

Soga, M., & Gaston, K. J. (2018). Shifting baseline syndrome: Causes, consequences, and implications. *Frontiers in Ecology and the Environment*, *0*(0). https://doi.org/10.1002/fee.1794

Southon, G. E., Jorgensen, A., Dunnett, N., Hoyle, H., & Evans, K. L. (2018). Perceived species-richness in urban green spaces: Cues, accuracy and well-being impacts. *Landscape and Urban Planning*, *172*, 1–10. https://doi.org/10.1016/j.landurbplan.2017.12.002

Sweet, F. S. T., Noack, P., Hauck, T. E., & Weisser, W. W. (2020). *The relationship between knowing and liking for 91 urban animal species among students*. SocArXiv. https://doi.org/10.31235/osf.io/gv6yx

Turner, W. R., Nakamura, T., & Dinetti, M. (2004). Global Urbanization and the Separation of Humans from Nature. *BioScience*, *54*(6), 585. https://doi.org/10.1641/0006-3568(2004)054[0585:GUATSO]2.0.CO;2

United Nations. (2018). *World Urbanization Prospects* (No. 2018/1; Population Facts, p. 2). United Nations - Department of Economic and Social Affairs. https://population.un.org/wup/

Wolf, L. J., Zu Ermgassen, S., Balmford, A., White, M. P., & Weinstein, N. (2017). Is Variety the Spice of Life? An Experimental Investigation into the Effects of Species Richness on Self-Reported Mental Well-Being. *PLOS ONE*, *12*(1), e0170225. https://doi.org/10.1371/journal.pone.0170225

Wood, E., Harsant, A., Dallimer, M., Cronin de Chavez, A., McEachan, R., & Hassall, C. (2018). Not all green space is created equal: Biodiversity predicts psychological restorative benefits from urban green space. *Frontiers in Psychology*, *9*. https://doi.org/10.3389/fpsyg.2018.02320

**Tables**:

**Table 1**: Summary statistics of the three generalized linear models exploring the relationships between perceived bird, butterfly and plant richness, and actual richness of the three taxa, nature relatedness, ecological knowledge, garden size and demographic variables (age, income, education, gender and urbanization level of childhood environment). Only variables that remained after model selection are presented. Some variables were not significant but strongly contributed to the model (AIC increased if variable removed). Estimates ± standard error are presented. Significance levels are provided: (·) p<0.1, (*) p<0.05, (**) p<0.01, (***) p<0.001.

|  | **Perceived bird richness** | **Perceived butterfly richness** | **Perceived plant richness** |
|---|---|---|---|
| Intercept | 0.66±0.16*** | 1.56±0.59** | **1.11±0.12*** |
| Nature relatedness | 0.04±0.04 | -0.23±0.15 | **0.06±0.02*** |
| Garden size |  | 0.10±0.04* | **0.09±0.03**** |
| Actual Richness (birds/butterflies/plants) |  | -0.17±0.10 |  |
| Income | 0.01±0.02 | -0.02±0.02 | **0.007±0.01** |
| Nature relatedness * richness |  | 0.05±0.02· |  |

Table 2: Determinants of well-being (Attention restoration; Attachment; Sense of identity and continuity with the past) for different taxa: plants, birds and butterflies, or spatial attributes: land cover. NR stands for nature relatedness, PR for perceived richness, A for abundance, SI for species identification skills, and WC for woody cover. Only variables that remained after model selection are presented. Some variables were not significant but strongly contributed to the model (AIC increased if variable removed). Estimates ± standard error are presented. Significance levels are provided: (·) p<0.1, (*) p<0.05, (**) p<0.01, (***) p<0.001.

| Taxa | Variable | Attention restoration | Attachment | Sense of identity & Continuity with past |
|---|---|---|---|---|
| PLANTS | Intercept | 11.64±2.11*** | 32.39±3.23*** | 17.35±3.38*** |
| | Garden size | 0.99±0.20*** | 0.65±0.30* | |
| | Nature relatedness (NR) | -0.64±0.50 | -0.66±0.78 | -0.24±0.82 |
| | Perceived richness (PR) | -1.40±0.40*** | -0.76±0.61 | -0.93±0.64 |
| | Species identification | -0.14±0.07· | | |
| | Age | 0.04±0.00*** | | 0.08±0.01*** |
| | Income | 0.45±0.12*** | -0.13±0.18 | 0.60±0.19** |
| | Gender (female) | - | | |
| | Gender (male) | 0.42±0.25· | | |
| | Education | -0.03±0.04 | 0.03±0.07 | -0.18±0.07* |
| | NR * PR | 0.41±0.10*** | 0.27±0.15· | 0.31±0.16· |
| BUTTERFLIES | Intercept | 9.72±2.01*** | 29.33±1.86*** | 19.41±2.90*** |
| | Garden size | 1.06±0.20**** | 0.65±0.33· | |
| | Nature relatedness (NR) | 0.06±0.07 | 0.62±0.29* | -0.09±0.60 |
| | Perceived richness (PR) | -1.73±0.71* | -0.17±0.29 | -0.27±0.30 |
| | Abundance (A) | | 0.02±0.05 | -0.77±0.29** |
| | Species identification (SI) | -0.13±0.07· | -0.71±0.29* | -0.80±0.30** |
| | Age | 0.04±0.00*** | | 0.08±0.01*** |
| | Income | 0.41±0.12*** | -0.14±0.18 | 0.58±0.19** |
| | Education | -0.06±0.05 | 0.02±0.07 | -0.20±0.08* |
| | NR * PR | 0.52±0.18** | | |
| | SI * PR | | 0.26±0.11* | 0.25±0.11* |
| | NR * A | | | 0.20±0.07** |
| | Intercept | 3.24±1.75· | 31.85±1.99*** | 11.07±2.92*** |
| | Garden size | 1.13±0.22*** | 1.40±0.35*** | 1.11±0.36** |
| | Nature relatedness (NR) | 1.80±0.0.38*** | 0.65±0.28* | 2.60±0.58*** |
| | Perceived richness | 0.22±0.16 | | 0.23±0.26 |
| BIRDS | Observed richness |  | -0.26±0.07*** | -0.32±0.07*** |
|  | Abundance (A) | 0.04±0.02 | -0.01±0.01 | 0.10±0.04* |
|  | Species identification (SI) | -0.13±0.07· | 0.25±0.21 |  |
|  | Age | 0.04±0.00*** |  | 0.08±0.01*** |
|  | Income | 0.41±0.12*** | 0.01±0.18 | 0.71±0.19*** |
|  | Education | -0.05±0.05 | -0.00±0.07 | -0.21±0.07** |
|  | SI * A |  | -0.00±0.00· |  |
|  | NR * A | -0.01±0.00· |  | -0.03±0.01** |
| LAND COVER | Intercept | 7.35±1.45 | 30.79±2.15*** | 17.08±2.38*** |
|  | Garden size | 1.06±0.26 | 0.85±0.40* |  |
|  | Nature relatedness (NR) | 0.71±0.30* | 0.03±0.46 | 0.48±0.48 |
|  | Woody cover (WC) | -4.29±2.30· | -6.12±3.50· | -6.41±3.60· |
|  | SDHI |  |  | -1.33±0.66* |
|  | Age | 0.04±0.00*** |  | 0.08±0.01*** |
|  | Income | 0.44±0.12*** | -0.10±0.18 | 0.62±0.19** |
|  | Education | -0.06±0.04 | 0.03±0.07 | -0.17±0.07* |
|  | NR * WC | 1.23±0.59* | 1.58±0.91· | 2.03±0.95* |

**Figures:**

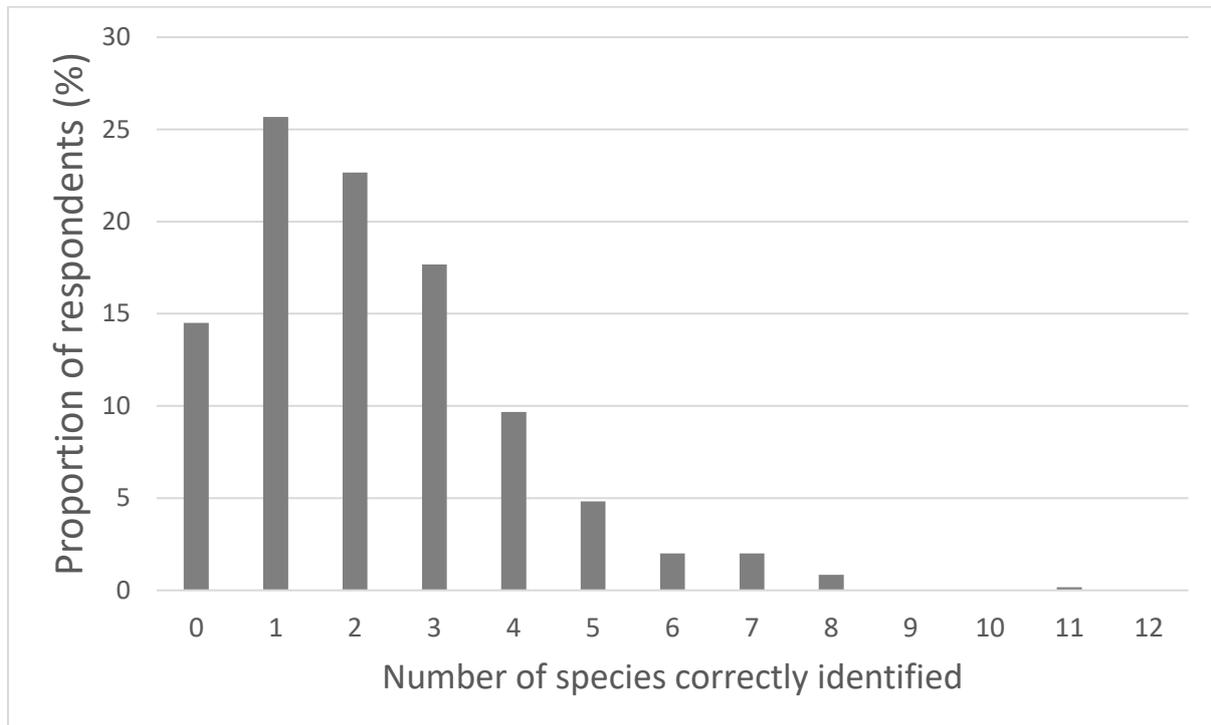

**Figure 1**: Proportion (%) of respondents who correctly identified species (species identification skills, 0 to 12 species).

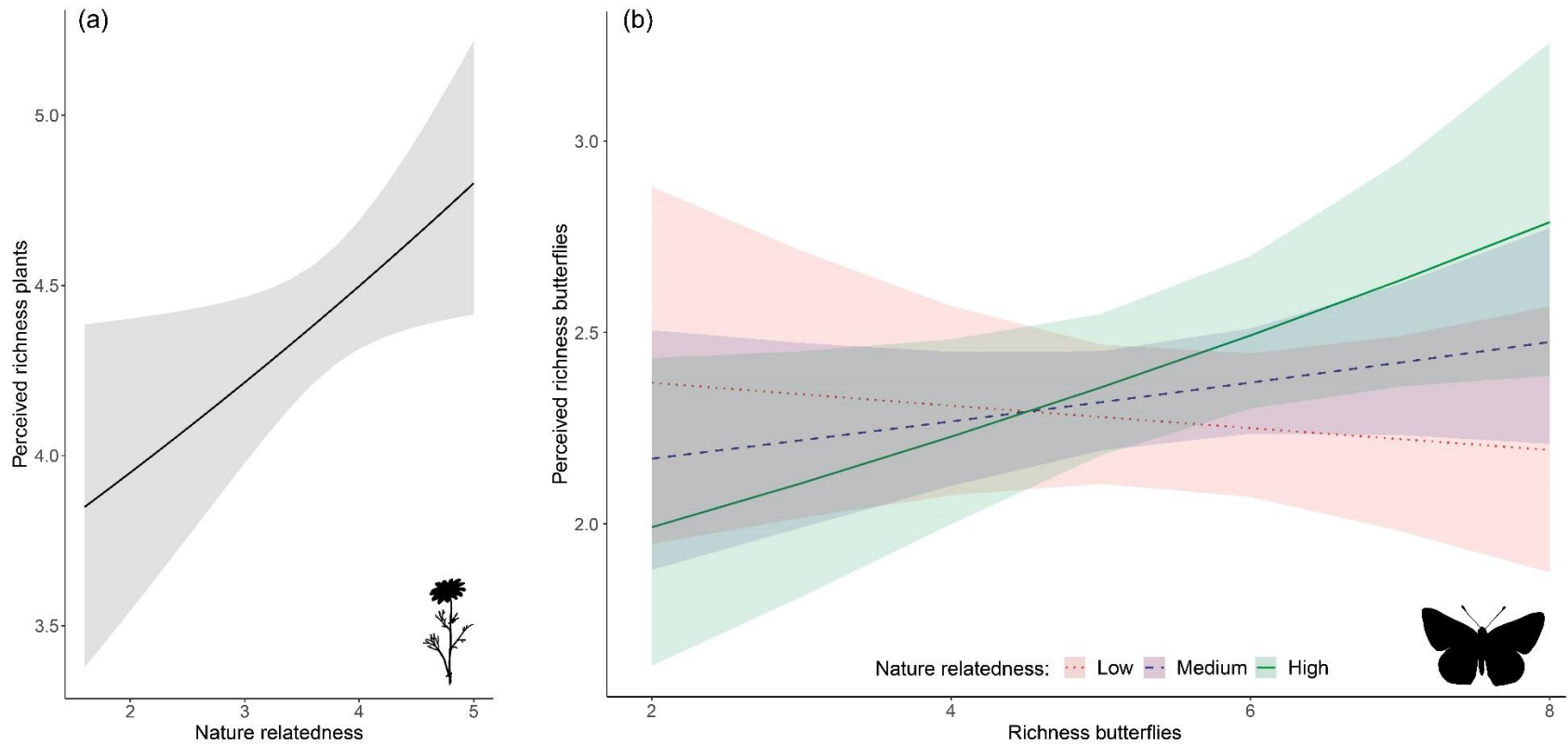

**Figure 2**: (a) Relationship between perceived richness of plants and nature relatedness (95% confidence interval). (b) Relationship between perceived richness of butterflies and actual richness of butterflies for varying levels of nature relatedness (95% confidence intervals). Levels of nature relatedness are displayed with different line types and color: low (dotted red), medium (dashed blue) and high (solid green).

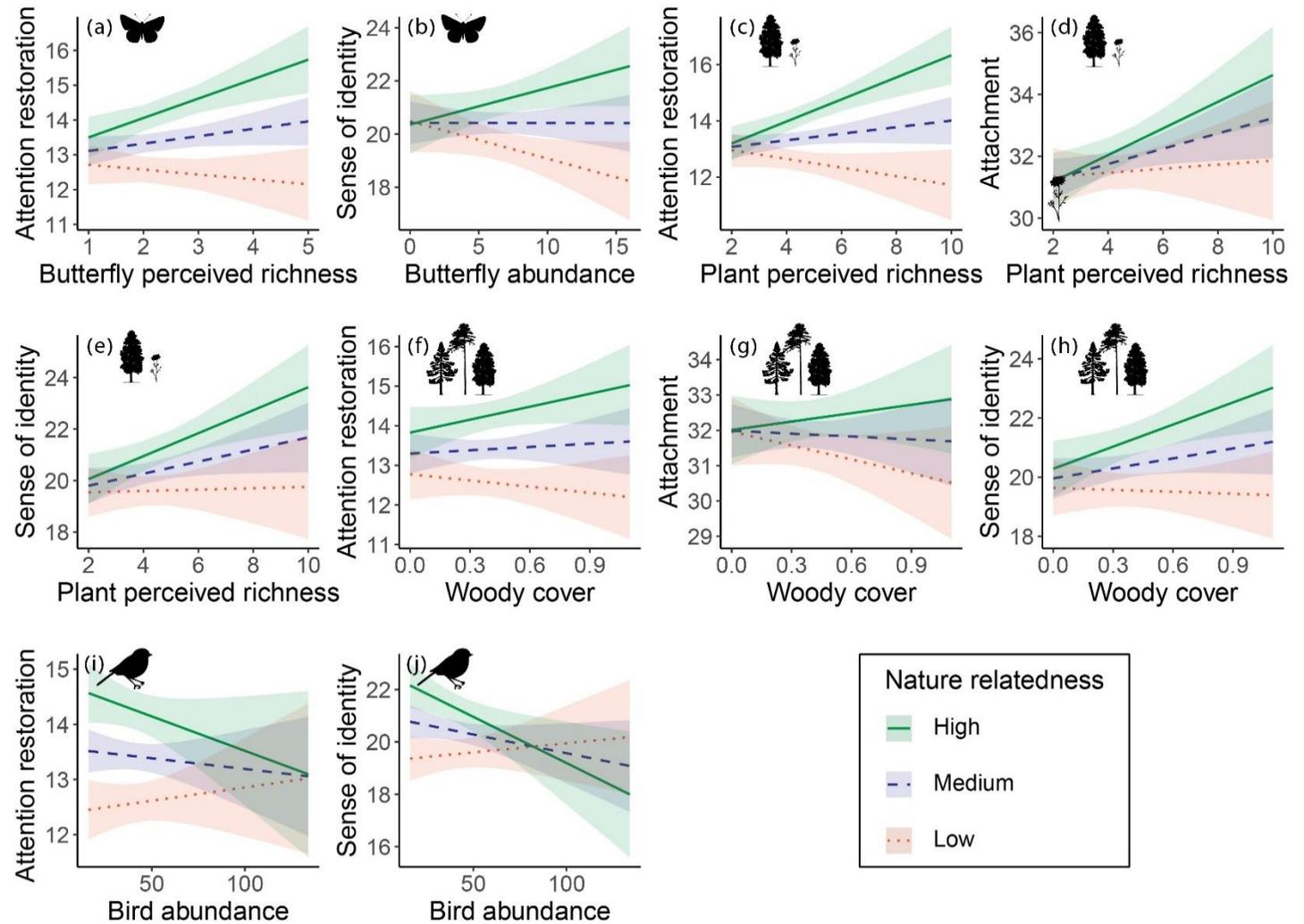

**Figure 3**: Relationship between different components of well-being (*attention restoration*, *attachment* and *sense of identity*) and *perceived richness* for plants and butterflies (a, c, d, e), *abundance* of butterflies and birds (b, i, j) or *land cover* (f, g, h) for varying levels of *nature relatedness* (95% confidence intervals). Levels of *nature relatedness* are displayed with different line types and color: low (dotted red), medium (dashed blue) and high (solid green).

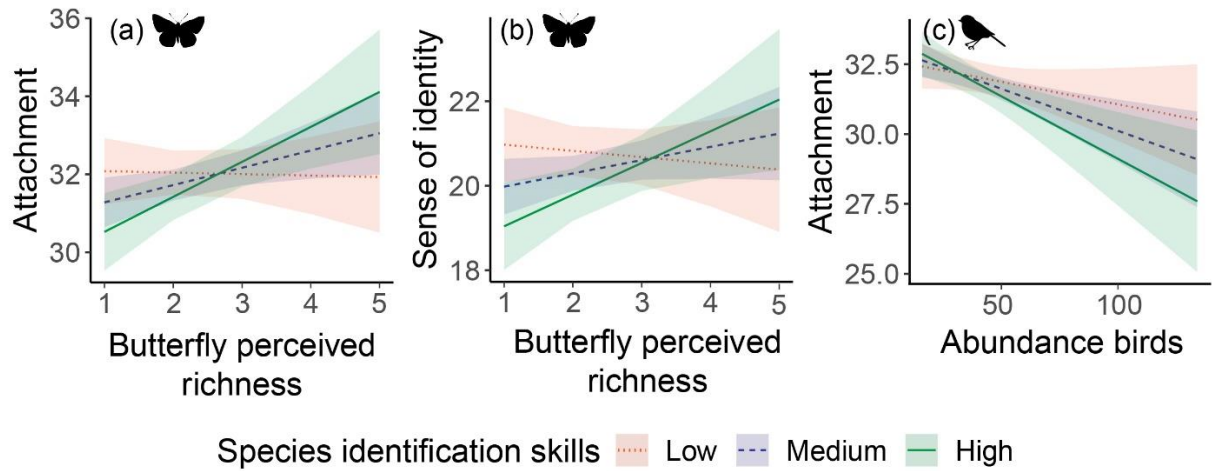

**Figure 4**: Relationship between different components of well-being (*attachment* and *sense of identity*) and perceived richness for butterflies (a, b), or abundance of birds (c) with varying levels of *species identification skills* (95% confidence intervals). Levels of *species identification skills* are displayed with different line types and color: low (dotted red), medium (dashed blue) and high (solid green).

**List of appendices**

Figure A1: Species presented to the respondents with respective numbers of respondents who indicated the correct genus name.
Table A1: Items loadings for well-being components and their internal validity (Cronbach's alpha), mean value and standard error.

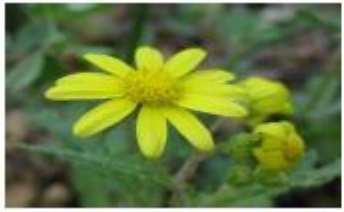 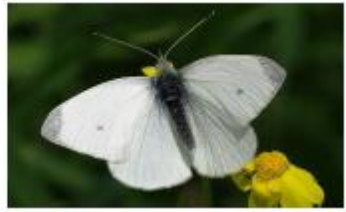 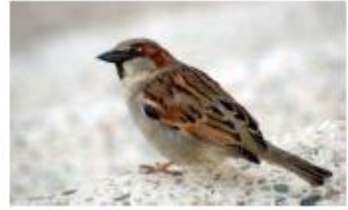
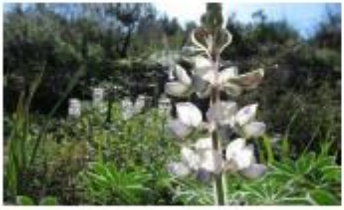 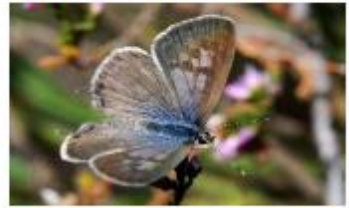 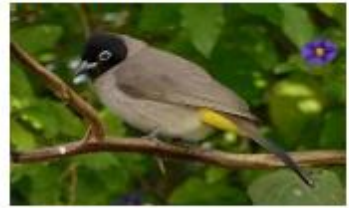
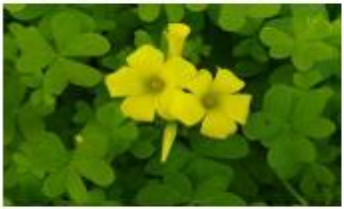 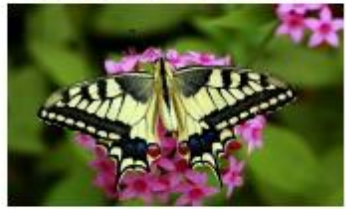 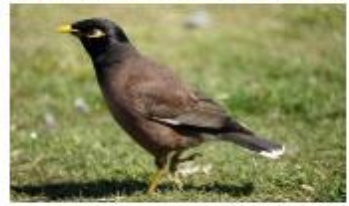
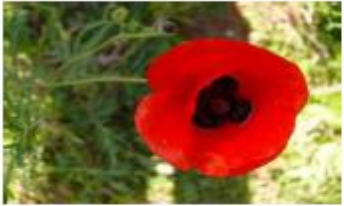 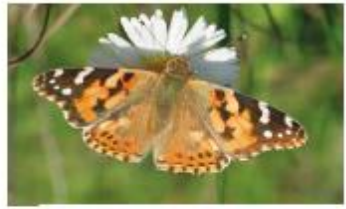 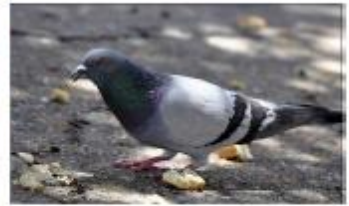

Senecio joppensis (109)  Pieris brassicae (32)  Passer domesticus (236)
Lupinus palaestinus (73)  Leptotes pirithous (1)  Pycnonotus xanthopygos (66)
Oxalis pes-caprae (91)  Papilio machaon (6)  Acridotheres tristis (38)
Papaver umbonatum (192)  Vanessa cardui (4)  Columba livia domestica (480)

Figure A1: Species presented to the respondents with respective numbers of respondents who indicated the correct genus name.

Table A1: Items loadings for well-being components and their internal validity (Cronbach's alpha), mean value and standard error.

| Dimensions of subjective well-being | Loadings[α] | Alpha | Mean±SE |
|---|---|---|---|
| **Attention restoration** | | | |
| Coming here clears my head | 0.622 | | |
| Being here makes me feel more connected to nature | 0.694 | 0.71 | 3.35±0.033 |
| I can easily think about personal matters when here | 0.648 | | |
| I gain perspective on life when I come here | 0.728 | | |
| **Attachment** | | | |
| I gain pleasure from using this park | 0.674 | | |
| I feel happy when I am in this park | 0.627 | | |
| I look forward to coming to this park in the future | 0.439 | | |
| Compared with other local parks, this park has many advantages | 0.429 | 0.83 | 3.99±0.023 |
| This park is well known as a desirable place to visit | 0.704 | | |
| I like this park | 0.690 | | |
| I find it easy to get familiar with this park | 0.577 | | |
| I am not satisfied with this park * | 0.645 | | |
| **Sense of identity and continuity with the past** | | | |
| I can influence decisions made about this park (e.g., by the council) | 0.414 | | |
| Lots of things in this park remind me of past experiences | 0.414 | | |
| This park feels almost like a part of me | 0.706 | | |
| I really miss this park when I am away from it for a long time | 0.632 | 0.79 | 2.91±0.029 |
| When I am in this park, I feel strongly that I belong here | 0.544 | | |
| I've had a lot of memorable experiences in this park | 0.420 | | |
| This park reflects the type of person I am | 0.617 | | |

\* Scores for statements have been reversed in analysis.
[α] Items that did not load: for restoration – "I find it hard to concentrate on difficult activities after being in the park", for sense of place" – "I don't feel at home in this park"; "I am proud of this park; It is difficult to find my way around the park